# Gap plasmon resonator arrays for unidirectional launching and shaping of surface plasmon polaritons

Zeyu Lei and Tian Yang[a]

*University of Michigan - Shanghai Jiao Tong University Joint Institute, State Key Laboratory of Advanced Optical Communication Systems and Networks, Key Laboratory for Thin Film and Microfabrication of the Ministry of Education, Shanghai Jiao Tong University, Shanghai 200240, China*

We report the design and experimental realization of a kind of miniaturized devices for efficient unidirectional launching and shaping of surface plasmon polaritons (SPPs). Each device consists of an array of evenly spaced gap plasmon resonators with varying dimensions. Particle swarm optimization is used to achieve a theoretical two dimensional launching efficiency of about 51%, under the normal illumination of a 5-μm waist Gaussian beam at 780 nm. By modifying the wavefront of the SPPs, unidirectional SPPs with focused, Bessel and Airy profiles are launched and imaged with leakage radiation microscopy.

Surface plasmon polaritons (SPPs) have been employed as an important method for confining light at subwavelength dimensions.[1–3] One critical issue in the development of plasmonic devices is the efficient and unidirectional launching of SPPs, which can't directly couple to free-space propagating lightwaves due to phase mismatch. Most SPP launching experiments require planewave illumination, which include prism coupling and grating coupling,[4] the latter of which includes both periodic[5] and aperiodic gratings[6]. Recently, a few efforts have been reported to minimize the launching device by coupling SPPs with focused beams. For example, an efficient unidirectional launcher based on aperiodic grooves with varying depths was experimentally demonstrated by Baron et al.[7] Another unidirectional launcher based on gap plasmon resonator (GPR) phase gradient metasurface was demonstrated by Pors et al.[8] Besides unidirectional coupling, beam shaping of SPPs is also an important issue. In recent years, SPPs with focused,[9,10] Bessel[11] and Airy[12,13] profiles have been experimentally demonstrated. However, all of the SPP beam shaping demonstrations have employed periodic grating coupling, with bi-directional radiation.

In this paper, we report the design and experimental realization of a kind of miniaturized devices for efficient unidirectional launching and shaping of SPPs. Each device consists of an aperiodic array of evenly spaced GPRs of varying dimensions.[8,14] At the free-space wavelength 780 nm and under the normal illumination of a focused 5-μm waist Gaussian beam, the theoretical two-dimensional coupling efficiency of our device is about 51% and the extinction ratio between the power of SPPs propagating to opposite directions is about 25:1. Further, by modifying

---

[a] **To whom correspondence should be addressed. Electronic mail: tianyang@sjtu.edu.cn.**

SPPs' wavefront based on the same aperiodic array of GPRs, unidirectional SPPs with focused, Bessel and Airy profiles are launched and imaged by leakage radiation microscopy (LRM).

The two-dimensional geometry of our devices is illustrated in Fig. 1(a). The top layer is an array of gold strips with varying widths and a uniform thickness of 30 nm. The middle layer is a continuous silicon dioxide spacer with a thickness of 30 nm. The bottom layer is an optically thick gold film. Each gold strip above the gold film comprise a GPR, whose resonant reflection phase can continuously change from 0 to $2\pi$ by varying its width. [8,14] The collective scattering of an array of GPRs couples free-space focused illumination to SPPs propagating in the region with no GPRs. The illumination is a transverse magnetic (TM) polarized, i.e., $x$-polarized, Gaussian beam with a 5-μm waist. The beam is at normal incidence onto, and with its focal plane on the gold strips. We use the finite-difference time-domain (FDTD) method (Lumerical FDTD Solutions) to calculate the unidirectional launching performance of the two-dimensional structure. The permittivity of gold at 780 nm was set to $-22.61+1.40i$ following Ref. 15, and that of silicon dioxide set to 2.11. A grid size of more than 100 points per wavelength (PPW) was used in the gaps and the nearby regions, which gradually decreased to about 22 PPW in the regions away from the GPRs. The simulation results were confirmed with a convergence test.

Particle swarm optimization (PSO) is an effective method to optimize more than three parameters simultaneously,[16] and has been used in our FDTD calculation to optimize our two-dimensional structure for maximum launching efficiency of SPPs along the $+x$ direction. The variable parameters include the width of each strip, the spacing $p$ between the nearest GPRs, and the position of the Gaussian beam relative to the coupling structure. The initial conditions and ranges of the variable parameters were set as follows. The numbers of GPRs were from 15 to 17. The strip widths are from 200 nm to 600 nm. $p$ was close to the SPP wavelength, which is 731 nm according to the insulator–insulator–metal (IIM) model.[17] The beam position was in the vicinity of the center of the structure. Usually a convergence was achieved after thousands of simulations. Subsequently, if the variable parameters converge to their boundary values, we modify the initial conditions and rerun the program. The optimization process has been repeated dozens of times to find the optimum configuration.

The optimized two dimensional structure is shown Fig. 1(a). It consists of 17 GPRs. The strip widths from right to left are 321, 159, 325, 334, 329, 321, 340, 334, 570, 547, 249, 283, 343, 349, 618, 633, and 677 nm. $p$= 741 nm.

The beam is centered 2.9 μm left to the rightmost strip. The beam is sitting off the center of the coupling structure in order to reduce the SPP propagation attenuation inside the coupling structure, and to suppress unwanted SPPs propagating towards the opposite direction. The $E_z$ profile of the scattered field is shown in Fig. 1(b). A 51% efficiency for launching SPPs into the +$x$ direction has been obtained by calculation, and also a 25:1 extinction ratio between SPPs launched into opposite directions. It is worth pointing out that the coupling performance considerably depends on the permittivity of metal. By setting the permittivity of gold to -21.20+0.74$i$, following Ref. 18, a higher launching efficiency of 60% is calculated.

It is interesting to notice that $p$ is slightly longer than the SPP wavelength, which is different from the commonly employed periodic grating coupling method. Within the coupling structure, both the free-space illumination and the already-launched SPPs scatter off the GPRs, but with different phases, then they interfere and couple to newly-launched SPPs in the designated propagation direction. $p$ is different from the SPP wavelength in order to compensate for the interaction between SPPs and GPRs. At the same time, under a finite-sized-waist beam illumination, the ratio between free-space illumination and already-launched SPPs that scatter off the same GPR is different for different GPRs, and consequently each GPR has a unique scattering phase in order to let the interference result, i.e. the newly launched SPPs, have the optimum phase.

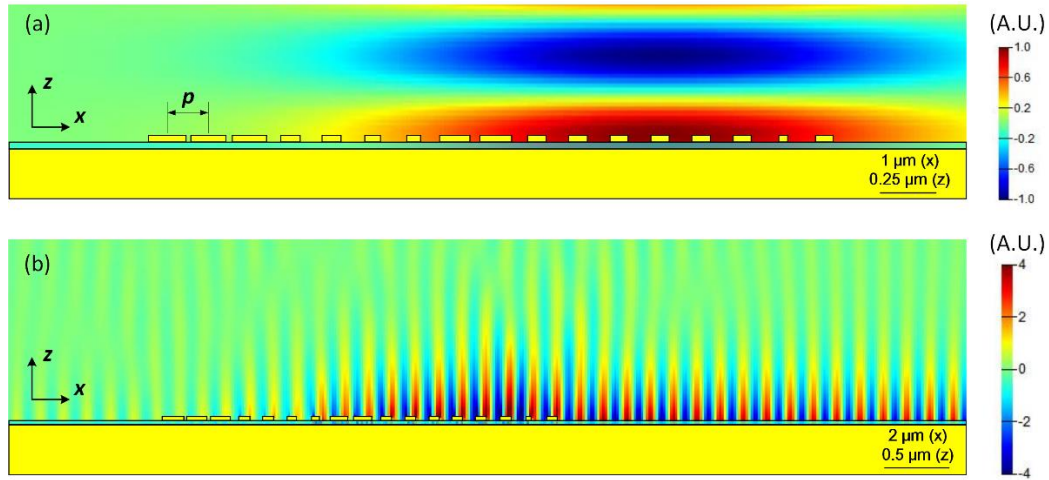

FIG. 1. (a) Electric field of the TM polarized incident Gaussian beam with the schematic of the aperiodic unidirectional SPP coupler. The structure is composed of 17 evenly spaced GPRs with period $p$=741 nm. The gold strips and the silicon dioxide spacer have the same thickness of 30 nm. The device is illuminated by a focused 5-μm waist Gaussian beam at 780 nm. The beam is centered 2.9 μm left to the rightmost strip. (b) $E_z$ profile of the excited SPPs calculated by two-dimensional FDTD method. $x$ and $z$ axes are not drawn to scale as labeled on the lower right corners.

It is also worth making a comparison between our structure and the GPR phase gradient metasurface for unidirectional SPP launching[8]. GPR is well-known for its ability to sweep the phase of free-space reflection from 0 to $2\pi$ by simply varying its transverse dimensions.[14] This full range phase sweep comes from the interference between GPR scattering and metal film reflection, which forms a Fano resonance. However, for SPP launching purpose, only GPR scattering is used to couple free-space illumination to SPPs, which by itself can't cover a phase range over $\pi$. Consequently, half of the metasurface should better contain no GPRs, which would otherwise destructively interfere with the SPPs launched by the GPRs in the other half of the metasurface. In addition, having fewer GPRs is helpful to reduce the coupling structure's internal SPP propagation loss. In fact, by following the GPR metasurface design method in Ref. 8, we were not able to obtain a launching efficiency higher than 30% under a 5-μm waist Gaussian beam at 780 nm using FDTD simulation when the thickness of the gold strips and silicon dioxide spacer are 30 nm.

We fabricated the proposed device and characterized its performance using leakage radiation microscopy (LRM).[19] First, a 50-nm-thick gold film was deposited onto a 200-μm-thick quartz substrate by electron beam evaporation, with a 3-nm-thick titanium film as an adhesion layer. Then a 30-nm-thick silicon dioxide spacer was deposited onto the gold film by magnetron sputtering. Lastly, the gold strips were fabricated on top of the silicon dioxide spacer by standard electron beam lithography and lift off techniques. The length of the strips is 14 μm. The scanning electron microscopy (SEM) image of the device is shown in Fig. 2(a). In the optical experiment, the collimated and TM-polarized Gaussian beam of a 780 nm laser was focused onto the device. The focusing objective was selected to obtain a focal spot with a radius about 5 μm. A 100× oil-immersion objective was used to collect the leakage radiation that was subsequently imaged by a monochrome CMOS camera. In this experiment, the 50 nm thick gold film provides enough SPP leakage radiation for LRM imaging, while it doesn't change the resonance and dispersion properties of the GPRs and SPPs significantly from that of an infinitely thick gold film. The LRM image is shown in Fig. 2(b), in which unidirectional launching of SPPs is well-illustrated. At 780 nm, the propagation distance of SPPs in the IIM model is calculated to be 21 μm, which agrees with the experimental observation.

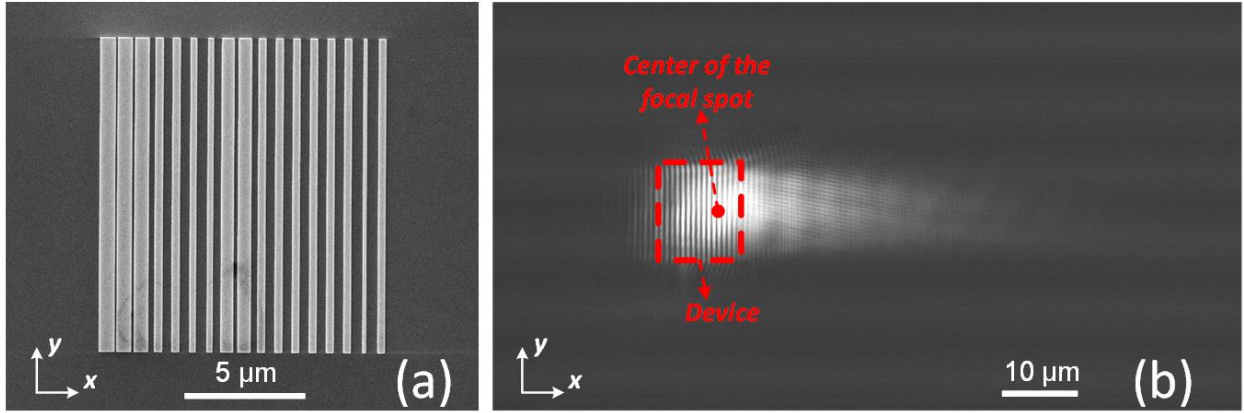

FIG. 2. (a) SEM image of a fabricated device for unidirectional SPP launching. (b) LRM image of the launched unidirectional SPPs. The focal spot is centered about 2.9 μm left to the right side of the device. The dashed box indicates the location of the device.

It is often desired to launch specifically shaped SPPs, such as focused, less diffracting or self-accelerating. Such beam shaping functions can be simply incorporated into our coupling structure by wavefront modification. This is achieved by reconfiguring the device along the additional dimension $y$. In the following we show the experimental results of launching focused, Bessel and Airy SPP beams.

The SEM images of the devices for launching focused SPPs are shown in Figs. 3(a) and 3(b). The LRM images are shown in Figs. 3(e) and 3(f). Here the strips are circularly curved to form a converging wavefront. Curvature radii of both 10 μm (Fig. 3(a)) and 25 μm (Fig. 3(b)) are shown. The SEM image of a device for launching a Bessel SPP beam is shown in Fig. 3(c). The LRM image is shown in Fig. 3(g). Here the strips, whose lengths are $L$, are folded to a specific angle, $\theta$, to produce an interference pattern that has the Bessel profile, just as what an axicon lens does in the three dimensional space[11]. The non-diffracting propagation distance of the Bessel beam is equal to the $x$-length of the interference region, which is approximately $L/\cos\theta$, if the Ohmic loss is ignored. The SEM image of a device for launching an Airy SPP beam is shown in Fig. 3(d). The LRM image is shown in Fig. 3(h). Here each strip is divided into 12 segments along the $y$ direction, with neighboring segments displaced along $x$ by half the SPP wavelength, i.e., 365.5 nm. The alternating 0 and $\pi$ phase profile of the coupled SPPs at the rightmost borderline of the device follows the Airy function[12]. Although the amplitude profile of the coupled SPPs is not modified to match the Airy function, the SPPs are still formed into an Airy beam since the other radiations diffract fast, as pointed out in Ref. 12. The amplitude profile mismatch results in less efficient coupling, and the parabolic trajectory in the LRM image is dimer compared to the Bessel beam even after optimizing the focal spot position in the experiment.

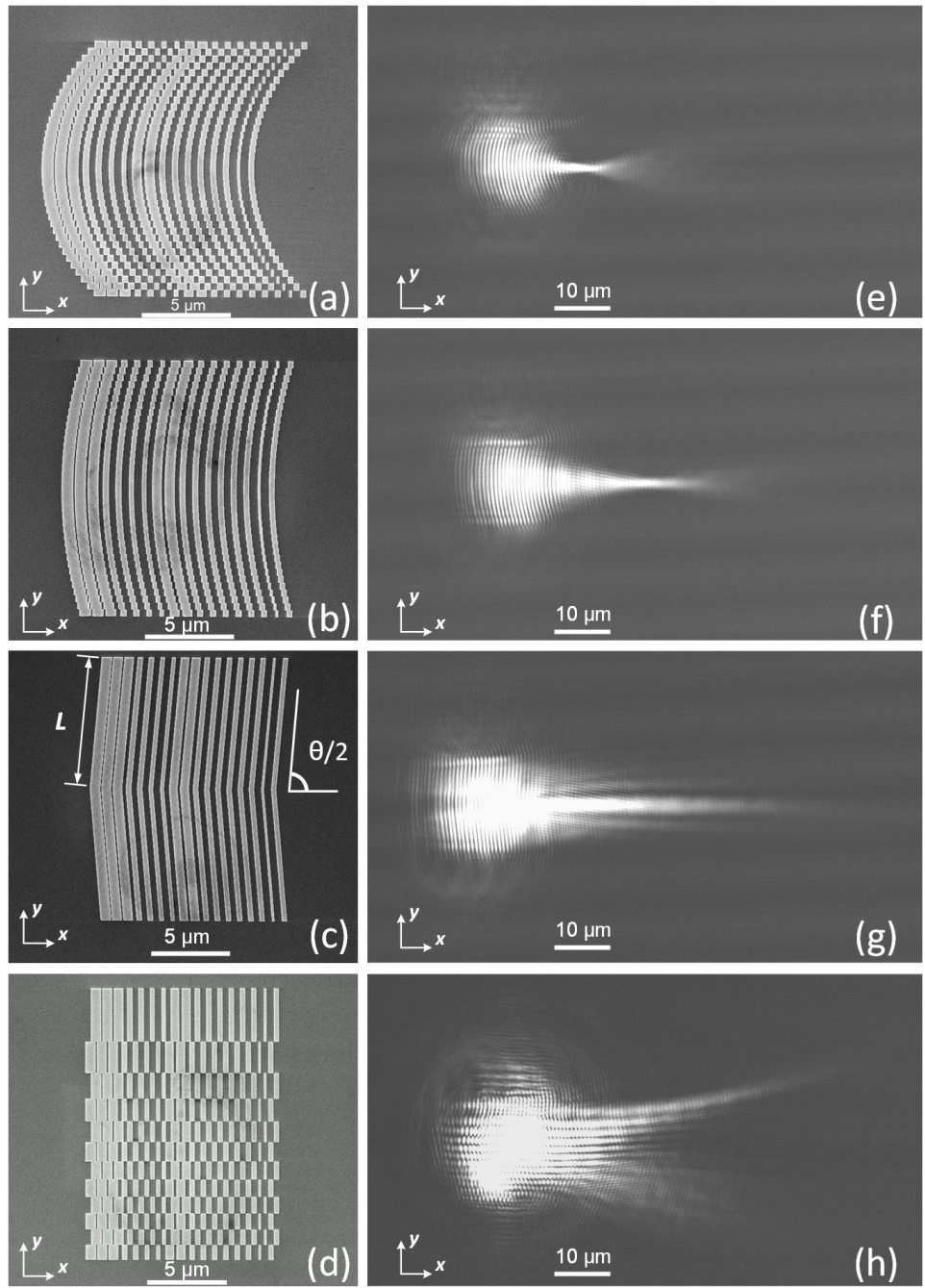

FIG. 3. SEM images of the fabricated devices for unidirectional launching of (a) focused SPP with focal length=10 μm, (b) focused SPP with focal length=25 μm, (c) Bessel SPP beam, the strips are folded to an angle of θ/2 = 85 °, and (d) Airy SPP beam. (e-h) show the corresponding LRM images of the launched SPPs from the devices whose SEM images are displayed in the same rows.

In summary, we have demonstrated a kind of devices for SPP launching under focused illumination. High efficiency unidirectional launching has been achieved by an array of evenly spaced GPRs, which resonantly modifies the local coupling phase and which has been optimized by the PSO method. Unidirectional SPP beams with focused, Bessel and Airy profiles have been launched by modifying the geometry of the GPR array to define the desired wavefront. Our method only requires a simple device fabrication procedure, and shows higher efficiency than previous GPR based metasurface devices at shorter wavelengths. These high efficiency, unidirectional and miniaturized SPP launching and shaping devices have a good potential for many applications in plasmonic integrated circuits and biosensing chips. [20,21]

## ACKNOWLEDGMENTS

This work is supported by the National Nature Science Foundation of China under Grant Nos. 61275168 and 11204177, and The National High Technology Research and Development Program of China (863 Program) under Grant 2015AA020944. Device fabrication is done at the Center for Advanced Electronic Materials and Devices of Shanghai Jiao Tong University, especially under the help from Jian Xu and Xuecheng Fu. We thank Prof. Tao Li and Qingqing Cheng at Nanjing University for help with experiments.

## REFERENCES

[1] E. Wijaya, C. Lenaerts, S. Maricot, J. Hastanin, S. Habraken, J. P. Vilcot, R. Boukherroub, and S. Szunerits, Curr. Opin. Solid. St. M. **15**, 208 (2011).
[2] M. S. Tame, K. R. McEnery, Ş. K. Özdemir, J. Lee, S. A. Maier, and M. S. Kim, Nat. Phys. **9**, 329 (2013).
[3] H. A. Atwater and A. Polman, Nat. Mater. **9**, 205 (2010).
[4] H. Raether, *Surface Plasmons* (Springer, Berlin, 1988).
[5] S. T. Koev, A. Agrawal, H. J. Lezec, and V. A. Aksyuk, Plasmonics **7**, 269 (2011).
[6] X. Huang and M. L. Brongersma, Nano. Lett. **13** (11), 5420 (2013).
[7] A. Baron, E. Devaux, J. C. Rodier, J. P. Hugonin, E. Rousseau, C. Genet, T. W. Ebbesen, and P. Lalanne, Nano. Lett. **11**, 4207 (2011).
[8] A. Pors, M. G. Nielsen, T. Bernardin, J. C. Weeber, and S. I. Bozhevolnyi, Light. Sci. Appl. **3**, e197 (2014).
[9] I. P. Radko, S. I. Bozhevolnyi, G. Brucoli, L. Martń-Moreno, F. J. Garcń-Vidal, and A. Boltasseva, Opt. Express **17**, 7228 (2009).
[10] L. Li, T. Li, S. Wang, S. Zhu, and X. Zhang, Nano. Lett. **11** (10), 4357 (2011).
[11] J. Lin, J. Dellinger, P. Genevet, B. Cluzel, F. de Fornel, and F. Capasso, Phys. Rev. Lett. **109**, 093904 (2012).
[12] A. Minovich, A. E. Klein, N. Janunts, T. Pertsch, D. N. Neshev, and Y. S. Kivshar, Phys. Rev. Lett. **107**, 116802 (2011).

[13]L. Li, T. Li, S. M. Wang, C. Zhang, and S. N. Zhu, Phys. Rev. Lett. **107**, 126804 (2011).
[14]S. Sun, K. Y. Yang, C. M. Wang, T. K. Juan, W. T. Chen, C. Y. Liao, Q. He, S. Xiao, W. T. Kung, G. Y. Guo, L. Zhou, and D. P. Tsai, Nano. Lett. **12**, 6223 (2012).
[15]P. B. Johnson and R. W. Christy, Phys. Rev. B **6**, 4370 (1972).
[16]J. Pond and M. Kawano, Proc. SPIE 7750, 775028 (2010)
[17]S. A. Maier, *Plasmonics Fundamentals and Applications*. (Springer, New York, 2007).
[18]*CRC Handbook of Chemistry and Physics*, edited by R. C. Weast (Chemical Rubber Corp., Boca Raton, FL, 1981).
[19]H. Ditlbacher, J. R. Krenn, A. Hohenau, A. Leitner, and F. R. Aussenegg, Appl. Phys. Lett. **83**, 3665 (2003).
[20]M. Righini, A. S. Zelenina, C. Girard, and R. Quidant, Nat. Phys. **3**, 477 (2007).
[21]T. W. Ebbesen, C. Genet, and S. I. Bozhevolnyi, Phys. Today **61**, 44 (2008).